\documentclass[aps,prb,twocolumn,showpacs,twoside,floatfix,superscriptaddress]{revtex4}

\usepackage{graphicx}
\usepackage{graphics}
\usepackage{amsmath}
\usepackage{amssymb}
\usepackage{mathrsfs}
\usepackage{epsf}

\begin{document}

\title{Multiplicative cross-correlated noise induced escape
rate from a metastable state}

\author{Jyotipratim Ray Chaudhuri}
\email{jprc_8@yahoo.com} \affiliation{Department of Physics, Katwa
College, Katwa, Burdwan 713130, India}

\author{Sudip Chattopadhyay}
\email{sudip_chattopadhyay@rediffmail.com} \affiliation{Department
of Chemistry, Bengal Engineering and Science University, Shibpur,
Howrah 711103, India}

\author{Suman Kumar Banik}
\email{skbanik@vt.edu} \altaffiliation{Present Address: Department
of Biological Sciences, Virginia Polytechnic Institute and State
University, Blacksburg, VA 24061-0406, USA.} \affiliation{Department
of Physics, Virginia Polytechnic Institute and State University,
Blacksburg, VA 24061-0435, USA}

\date{\today}

\begin{abstract}
We present an analytical framework to study the escape rate from a
metastable state under the influence of two external multiplicative
cross-correlated noise processes. Starting from a phenomenological
stationary Langevin description with multiplicative noise processes,
we have investigated the Kramers' theory for activated rate
processes in a nonequilibrium open system (one-dimensional in
nature) driven by two external cross-correlated noise processes
which are Gaussian, stationary and delta correlated. Based on the
Fokker-Planck description in phase space, we then derive the escape
rate from a metastable state in the moderate to large friction limit
to study the effect of degree of correlation on the same. By
employing numerical simulation in the presence of external
cross-correlated additive and multiplicative noises we check the
validity of our analytical formalism for constant dissipation, which
shows a satisfactory agreement between both the approaches for the
specific choice of noise processes. It is evident both from
analytical development and the corresponding numerical simulation
that the enhancement of rate is possible by increasing the degree of
correlation of the external fluctuations.
\end{abstract}

\pacs{05.40.-a, 02.50.Ey, 82.20.Uv}

\maketitle


\section{Introduction}

Rate theory deals with the passage of a system from one stable local
minima of the energy landscape to another via a potential energy
barrier providing relevant information on the long-time behavior of
the system with different metastable states and hence is a very
useful and pedagogical model for the microscopic description and
understanding of a wide range of physical, chemical and biological
phenomena. Examples include diffusion of atoms in solids or on
surfaces, isomerization reactions in solution, electron transfer
processes, and ligand binding in proteins or protein folding
reactions. Not only that the phenomena of flux transitions in
superconducting quantum interference devices can also be modeled by
rate theory. Kramers \cite{ref1} pointed out that the passage of a
system from a metastable state due to thermal agitation can be
considered as a model Brownian particle trapped in a one-dimensional
well representing the reactant state which is separated by a barrier
of finite height from a deeper well signifying the product state.
The particle is supposed to be immersed in a medium such that the
medium exerts damping force on the particle but at the same time
thermally activates it so that the particle may gain enough energy
from its surroundings to cross the barrier. Thus, the surrounding
fluid provides friction and random noise. Conventionally, the
friction is assumed to be Markovian and the noise, Gaussian and
white. Kramers theory provides a useful framework for the
explanation and prediction of rates and mechanisms for various
barrier crossing phenomena and for chemical reactions in particular.
\cite{simon,diau,mccann,ref2,ref3,ref4,nitzan,grotehynes,pollakjcp85}

Most of the above mentioned works are basically dealt either with
phenomenological Langevin equation or by microscopic
system-reservoir formulation with both bi-linear or non-linear
coupling scheme. Although, the effect of system-bath linear coupling
and associate consequences in different branches of molecular
sciences is well studied, the modeling and the proper microscopic
interpretation of the nature of non-linear coupling and
corresponding manifestation on the various molecular phenomena is a
very challenging task posed in front of researchers in this field.
Tanimura and co-workers \cite{tanimura} explained the phenomena of
elastic and inelastic relaxation mechanism and their cross effect in
vibrational and Raman spectroscopy using non-linear system-bath
model.

It is now a well documented fact that solvent plays a very crucial
role in computing the dynamics of a Brownian particle.
\cite{res,jrc,rig,eli,moi,popov,hsia,jpa,jcp} The dynamics of a
model Brownian particle in an inhomogeneous solvent generally leads
to state dependent diffusion due to the appearance of multiplicative
noise and state dependent dissipation. \cite{jpa,jcp,jrc4} Although
modeling of Brownian motion in presence of inhomogeneous solvent is
a nontrivial task, several approaches have been made to study
different aspects of molecular sciences, e.g., activated rate
processes, \cite{ref2,jrc4,pollakrev,pol} noise induced transport,
\cite{jpa,jcp,dref7,dref10,dref12} stochastic resonance \cite{mar}
and laser and optics. \cite{dref6}

In the overwhelming majority of the above mentioned treatments,
fluctuations experienced by the system of interest is of internal
origin so that the fluctuations and dissipation get related through
the celebrated fluctuation-dissipation relation (FDR). \cite{ref5}
However, in some situations, the physical origin of dissipation and
of fluctuations are different as well as independent.
\cite{jpa,jcp,jrc4,ref6,huang,jrc1,jrc2,jrc3,bravo} Thus, there
exists no balance between the influx (efflux) of energy to (from)
the system and hence there exist no FDR. As a result of this, such
systems (usually termed as nonequilibrium open system
\cite{lindenberg}) do not approach thermal equilibrium
asymptotically and many interesting phenomena are observed thereof.
\cite{ref6,huang}

To the best of our knowledge, creation of an open system in a
nonequilibrium process can be realized in many ways. To mention, we
briefly discuss about few of them in the following which are
relevant to the present work. An additional perturbation applied to
the heat bath can excite few bath modes which effectively leads to
the creation of a nonstationary process where the nonstationarity
mainly gets reflected in the dissipation kernel.
\cite{jrc,rig,popov,millonas} Dynamics in this case is mainly guided
by an irreversible process. \cite{rig,popov} An extensive analysis
of this approach can be found in the very recent work of Popov and
Hernandez. \cite{popov} External perturbation on the other hand can
directly excite the system mode itself (leaving the bath mode alone)
or the local bath modes (leaving the system mode alone). In both the
situations the additional energy input by an independent source
leads to a shift in the temperature, thus creating an effective
temperature like quantity, \cite{jpa,jcp,jrc4,jrc1,jrc2,jrc3,jrc5}
keeping the underlying dynamical process stationary. Another
immediate effect of the breakdown of FDR in this case is the
creation of a steady state instead of an equilibrium state in the
long time limit. To study the escape rate from a metastable state
under the influence of external cross-correlated noise processes we
have adopted one of the situations mentioned in the latter case to
create an open system. In the present work we drive the reaction
coordinate (the system mode) by multiplicative cross-correlated
noise processes, leaving the bath modes as it is. In such a
situation, the Brownian particle is energized by an extra input of
energy, in addition to the thermal energy provided by the heat bath.
This leads to, what we have tried to study in this paper, an
enhancement of barrier crossing event in the activated rate
processes within the framework of Markovian stationary dynamics.

The barrier crossing dynamics with multiplicative and additive white
noise processes aroused strong interest in the early eighties, where
noise forces that are present simultaneously in the dynamical system
were usually treated as random variables uncorrelated with each
other. However, there are situations where fluctuations in some
stochastic process may have common origin. If this happens then the
statistical properties of the fluctuations should not be very much
different and can be correlated to each other. Nonlinear stochastic
systems including noise terms have drawn interest on a wide scale
owing to their numerous applications in the field of molecular
sciences. Generally it is observed that under such a situation, the
noise affects the dynamics through a system variable. The
cross-correlated noise processes were first considered by Fedchenia
\cite{ref8} in the context of hydrodynamics of vortex flow where the
author introduced cross-correlation among the fluctuations from a
common origin that appear in the time evolution equation of
dimensionless modes of flow rates. The interference of additive and
multiplicative white noise processes in the kinetics of the bistable
systems was first considered by Fulinski and Telejko \cite{ref9}
where they mentioned the physical possibility of a cross-correlated
noise. Madureira {\it et al.} \cite{ref10} have pointed out the
probability of cross-correlated noise in a realistic model (ballast
resistor) showing bi-stable behavior  of the system. Recently, Mei
{\it et al.} \cite{ref11} have studied the effects of correlations
between additive and multiplicative noise on relaxation time in a
bi-stable system driven by cross-correlated noise.

It is now well accepted that the effect of correlation between
additive and multiplicative noise (which is also termed as
correlated noise processes) is considered indispensable in
explaining phenomena such as steady state properties of a single
mode laser, \cite{zhu} bistable kinetics, \cite{bikinetic}
stochastic resonance in linear system, \cite{sr} steady-state
entropy production, \cite{entropy} stochastic resonance \cite{ref12}
and transport of particles, \cite{li} etc. Zhu \cite{zhu}
investigated theoretically the statistical fluctuations of a single
mode laser that include correlations between additive and
multiplicative white noises and showed that the effect of
correlation can lead to larger intensity fluctuations. One can
utilize this intensity fluctuations to induce a narrow peak in the
probe absorption spectrum as well as significantly modify the
emission spectra of matter strongly resonant with laser field. As
shown by Ai {\it et al.} \cite{ai} one can use the logistic
differential equation to analyze the effects of environmental
fluctuations on cancer cell growth using correlated Gaussian white
noise scheme whereby the degree of correlation of the noise
(environmental intensive fluctuations) can cause tumor cell
extinction. Berdichevsky and Gitterman \cite{sr} showed that the
maxima of signal to noise ratio, as a function of the asymmetry of
noise, disappears in the absence of the coupling between additive
and multiplicative white noises. Very recently Ghosh \textit{et al.}
\cite{ghosh} have used multiplicative correlated noise processes to
study the splitting of Kramers' rate in a symmetric triple well
potential.

As the presence of the cross-correlated noise changes the Langevin
dynamics of the system, \cite{ref13,jrc5} it is expected that there
may exist some additional effect of cross-correlation on the escape
rate of a metastable state. Study of this additional effect can
perfectly serve as a motivation of our work presented in this paper.
To achieve this we extend our recently developed theoretical
approach \cite{jrc5} to study the effect of multiplicative
cross-correlated noise on the escape rate from a metastable state.
In this present study, the external fluctuation applied to the
system under consideration is assumed to be independent of the
system's characteristic dissipation. In this article, we study the
reaction rate (in one dimension) under the influence of both the
internal (thermal) noise and external (non-thermal) cross-correlated
noises simultaneously to examine the role of correlation between
external delta correlated fluctuations on the escape rate from a
metastable state when the dissipation is space dependent and the
noises appear multiplicatively in the dynamical equation. In our
model the system is externally driven by two cross-correlated
fluctuations. Starting from phenomenological Langevin equation with
space dependent dissipation and multiplicative noises, we construct
the corresponding Fokker-Planck equation in phase space. Under
proper boundary conditions, we solve the Fokker-Planck equation to
calculate the escape rate for moderate to large dissipation. In the
numerical implementation of our development the steady state rate
expression (derived analytically) is checked with stochastic
simulation to establish the fact that the incorporation of external
perturbation through cross-correlated fluctuations in to the
traditional Kramers' model enhances the rate across the barrier top.

The organization of the paper is as follows. In section II, starting
from a phenomenological Langevin equation for an open system
nonlinearly coupled with the environment and simultaneously acted
upon by two cross-correlated multiplicative white noise processes,
we construct the Fokker-Planck description of the underlying
stochastic process which is multiplicative in general. We then
calculate the escape rate from a metastable state to examine the
barrier crossing dynamics. A typical example has been considered in
section III to study the effect of the cross-correlated fluctuations
on the escape rate. The paper is then concluded in section IV. To
make the paper self contained we provide the derivation of the
Fokker-Planck equation and the calculation of escape rate in the two
appendices.


\section{Correlated noise induced escape from a metastable state}

We consider the motion of a particle of unit mass moving in a
Kramers type potential $V(x)$ such that it is acted upon by random
force $f(t)$ of internal origin, i.e., $f(t)$ originates due to the
coupling of the system with its environment and hence is connected
to the friction through the fluctuation-dissipation relation. Apart
from the internal noise, we assume that the system is acted upon by
two external Gaussian noise processes, $\epsilon(t)$ and $\pi(t)$,
both of which have a common origin and consequently are correlated.
Thus, from the very mode of description of our model, it is evident
that the system is open in nature as it is driven externally by two
correlated fluctuations. The dynamics of the particle is governed by
the Langevin equation
\begin{equation}\label{eq1}
\ddot{x}= - \Gamma(x) \dot{x} - V^{\prime}(x) + h(x)f(t) +
g_1(x)\epsilon(t) + g_2(x)\pi(t),
\end{equation}

\noindent with
\begin{equation}\label{eq2}
h(x) = \sqrt{k_B T \Gamma(x)},
\end{equation}
where $\Gamma(x)$ is the space dependent friction that arises due to
non-linear system-reservoir coupling. \cite{lindenberg} Here,
$g_1(x)$ and $g_2(x)$ are two arbitrary smooth functions of $x$ and
their presence makes the external noise processes multiplicative.
$T$ is the thermal equilibrium temperature and $k_B$ is the
Boltzmann constant. $\epsilon(t)$ and $\pi(t)$ are Gaussian white
noise processes with statistical properties
\begin{subequations}
\begin{eqnarray} \label{eq3a}
\langle \epsilon(t) \rangle & = & \langle \pi (t) \rangle = 0 \\
\label{eq3b} \langle \epsilon(t) \epsilon (t^{\prime}) \rangle & = &
2D_\epsilon \delta
(t-t^{\prime}) \\
\label{eq3c}
\langle \pi(t) \pi(t^{\prime})\rangle &=& 2 D_\pi \delta (t-t^{\prime}) \\
\label{eq3d} \langle \epsilon(t) \pi (t^{\prime}) \rangle &=&
\langle \pi(t) \epsilon (t^{\prime}) \rangle =2\lambda
\sqrt{D_\epsilon D_\pi} \delta (t-t^{\prime}).
\end{eqnarray}
\end{subequations}

\noindent In the above equations (\ref{eq3b}-\ref{eq3d}),
$D_\epsilon$ and $D_\pi$ are the strength of the fluctuations
$\epsilon(t)$ and $\pi(t)$, respectively and $\lambda$ $(0 \leqslant
\lambda < 1)$ denotes the degree of correlation between the noise
processes $\epsilon(t)$ and $\pi(t)$. The internal noise $f(t)$ is
also assumed to be Gaussian and delta correlated with statistical
properties
\begin{equation}\label{eq4}
\langle f(t) \rangle = 0 \text{ and } \langle f(t)f(t^{\prime})
\rangle = 2 \delta (t-t^{\prime}).
\end{equation}

\noindent In the above equations $\langle \cdots \rangle$ implies
the average over the realizations of the noise (external or
internal) processes. Eq.(\ref{eq2}) along with the second relation
of Eq.(\ref{eq4}) comprises the fluctuation-dissipation which
relates the damping function $\Gamma(x)$ with the fluctuation
$f(t)$. The external noise processes $\epsilon(t)$ and $\pi(t)$ are
independent of the dissipation function and there is no
corresponding fluctuation-dissipation relation. We further assume
that $f(t)$ is independent of $\epsilon(t)$ and $\pi(t)$ so that
\begin{equation}\label{eq5}
\langle f(t) \epsilon(t') \rangle = \langle f(t) \pi(t') \rangle =0.
\end{equation}

In the absence of the external noise processes, the system being
closed, the fluctuation-dissipation relation eventually brings it to
a stationary state and consequently one can examine the barrier
dynamics of the system using standard methods applicable for a
thermodynamically closed system. The external fluctuations and their
correlation modify the dynamics of activation in two ways. First,
they influence the dynamics in the region around the barrier top so
that the effective stationary flux across it gets modified. Second,
in the presence of these fluctuations, the equilibrium distribution
of the source well is disturbed so that one has to consider a new
stationary distribution, if any, instead of the standard equilibrium
Boltzmann distribution. This new stationary distribution must be a
solution of the Fokker-Planck equation around the bottom of the
source well region and serves as an appropriate boundary condition
analogous to Kramers problem.


Keeping this in mind we write the corresponding Fokker-Planck
equation in the phase space [i.e. in $(x,v)$ space of the system,
where $v=\dot{x} \equiv dx/dt$], describing the dynamics of the
Langevin equation (\ref{eq1}) [see Appendix-A for detailed
derivation]
\begin{eqnarray}
\frac{\partial P}{\partial t}& = & -v\frac{\partial P}{\partial
x}+[\Gamma(x)v+V'(x)]\frac{\partial P}{\partial v} +
A (x) \frac{\partial^2 P}{\partial v^2} \nonumber \\
&& + \Gamma(x)P, \label{eq24}
\end{eqnarray}

\noindent
where
\begin{eqnarray}
\label{eq25}
A(x) & = & k_B T \Gamma(x) + g^2(x) \\
\label{eq26} g(x) & = & \left \{ D_\epsilon g_1^2(x) + 2 \lambda
\sqrt{D_\epsilon D_\pi} g_1(x) g_2(x) \right. \nonumber \\
&& \left. + D_\pi g_2^2(x) \right \}^{1/2}.
\end{eqnarray}

\noindent The diffusion coefficient $A(x)$ in the Fokker-Planck
equation (\ref{eq24}) is an implicit function of the correlation
parameter $\lambda$ (see the expression of $g(x)$). Thus by
increasing the value of $\lambda$ one can increase the value of the
diffusion coefficient. This property gets directly reflected in the
final rate expression and in the expression of effective temperature
(see Eqs.(12) and (13)). In deriving Eq.(\ref{eq24}), we have made
use of Eqs.(\ref{eq3a}-\ref{eq3d}) and (\ref{eq4}) and the fact that
$f(t)$ is independent of $\epsilon(t)$ and $\pi(t)$. It should be
noted that when the noise is purely internal and for linear
system-reservoir coupling, Eq.(\ref{eq24}) reduces to the Kramers
equation. \cite{ref1}

Kramers' model for a chemical reaction consists of a particle
undergoing Brownian motion whose coordinate $x$ corresponds to the
reaction coordinate and $v=dx/dt$ the reaction rate. In addition to
that in Kramers' original treatment, \cite{ref1} the dynamics of the
Brownian particle was governed by Markovian random process. Since
the work of Kramers, a number of workers have extended Kramers model
for non-Markovian case and for state dependent diffusion to derive
the expression for the escape rate. \cite{ref2} In order to allow
ourselves a comparison with the Fokker-Planck equation of other
forms, we note that though the underlying dynamics is Markovian, the
diffusion coefficient in Eq.(\ref{eq26}) is coordinate dependent. It
is customary to get rid of this dependence by approximating the
coefficients at the barrier top or potential well where we need the
steady-state solution of Eq.(\ref{eq26}). One may also use mean
field solution of Eq.(\ref{eq26}) obtained by neglecting the
fluctuation terms and putting approximate stationary condition in
the diffusion coefficient.

For harmonic oscillator with frequency $\omega_0$, $V(x) \approx E_0
+ \omega_0^2 (x-x_0)^2/2$, the linearized version of the
Fokker-Planck equation can be represented as
\begin{eqnarray}\label{eq27}
\frac{\partial P}{\partial t} & = &  -v\frac{\partial P}{\partial x}
\nonumber + \Gamma (x_0) P +[\Gamma (x_0) v+\omega_0^2
(x-x_0)]\frac{\partial P}{\partial v} \nonumber \\
&& + A_0\frac{\partial^2 P}{\partial v^2},
\end{eqnarray}

\noindent where $A_0 = k_B T \Gamma(x_0) + g^2(x_0)$, is calculated
at the bottom of the potential ($x \approx x_0$). The general steady
state solution of Eq.(\ref{eq27}) becomes
\begin{equation}\label{eq28}
P_0^{st}(x,v) = \frac{1}{Z} \exp \left (- \frac{v^2}{2 D_0}-
\frac{\omega_0^2 (x-x_0)^2}{2 D_0} \right ),
\end{equation}

\noindent with $D_0=A_0/\Gamma(x_0)$ and $Z$ is the normalization
constant. The solution (\ref{eq28}) can be verified by direct
substitution in the steady state version of the Fokker-Planck
equation (\ref{eq27}), namely
\begin{eqnarray*}\label{eq30}
&& -v\frac{\partial P_0^{st}}{\partial x} \nonumber + \Gamma(x_0)
P_0^{st} +[\Gamma(x_0) v+\omega_0^2 (x-x_0)]\frac{\partial
P_0^{st}}{\partial v} \nonumber \\
&&+ A_0\frac{\partial^2 P_0^{st}}{\partial v^2} = 0.
\end{eqnarray*}

\noindent It should be noted that the distribution (\ref{eq28}) is
not an usual equilibrium distribution, rather it serves as
stationary distribution for the non-equilibrium open system. This
stationary distribution plays the role of an equilibrium
distribution of the closed system which can, however, be recovered
in the absence of the external noise.

We now embark on the problem of decay of a meta stable state in the
presence of external cross-correlated noise processes. In Kramers'
approach, the particle coordinate $x$ corresponds to the reaction
coordinate, and its values at the minima of the potential well
$V(x)$ are separated by a potential barrier to describe the reactant
and product states. Linearizing the motion around the barrier top at
$x \approx x_b$, the steady state version of the Fokker-Planck
equation corresponding to Eq.(\ref{eq24}) reads as
\begin{eqnarray}
&& -v\frac{\partial P_b^{st}}{\partial x} -\omega_b^2 (x - x_b)
\frac{\partial P_b^{st}}{\partial v} + \Gamma (x_b) \frac{\partial
(v P_b^{st})}{\partial v}
\nonumber \\
&& + A_b \frac{\partial^2 P_b^{st}}{\partial v^2}  = 0, \label{eq31}
\end{eqnarray}

\noindent where, $V(x) \approx E_b - \omega_b^2 (x - x_b)^2/2$ with
$\omega_b^2> 0$, and the suffix `$b$' indicates that coefficients
are to be calculated using the general definition of $A$ at the
barrier top.


After imposing appropriate physically motivated boundary conditions
we then derive the escape rate for nonequilibrium open systems valid
in the moderate to strong friction limit [see Appendix-B for
detailed calculation]
\begin{equation}\label{eq48}
k = \frac {\omega_0}{2\pi}
\frac{D_b}{\sqrt{D_0}}\sqrt{\frac{\Lambda}{1+ \Lambda D_b}} \exp
\left(-\frac {E}{D_b} \right),
\end{equation}

\noindent where $E=E_b-E_0$ is the potential barrier height, and
$D_b=A_b/\Gamma(x_b)$. Already we have mentioned in the
introduction, open system mean the system is not thermodynamically
closed as it is driven by two noises of external origin and
consequently there is no fluctuation-dissipation relation which is
mainly responsible to establish the thermal equilibrium (for closed
system). It is also important to note that the escape from a
metastable state intrinsically a non-equilibrium phenomenon as
fluctuations establish a steady mass motion to give rise to a
non-vanishing constant current over the potential barrier. In the
steady state rate expression (\ref{eq48}), all information about the
thermal temperature due to internal noise processes, the strength of
both the external noises and the position dependent dissipation,
$\Gamma (x)$, are hidden in the quantities $D_0$, $D_b$ and
$\Lambda$. Furthermore, the rate constant $k$ is an implicit
function of the degree of correlation, $\lambda$, between the
external noise processes. We conclude this section by mentioning the
fact that the $\lambda$-dependence of the effective temperature
($D_b$) of our model does not depend on the detailed forms of the
coupling functions $g_1(x)$ and $g_2(x)$.


\section{Analysis of the interference between two external fluctuations}

In this section we present the numerical implementation of our
presently developed theory to establish its potentiality and
applicability to demonstrate the effect of correlation of two
external noises $\epsilon (t)$ and $\pi (t)$ on the rate.
Eq.(\ref{eq48}) is the theoretical expression for the escape rate
and is applicable for any two functions $g_1(x)$ and $g_2(x)$
appeared in equation (\ref{eq1}). To test the validity of the rate
expression (\ref{eq48}), one needs to assume the particular forms of
$g_1(x)$ and $g_2(x)$. In addition to that an explicit form of the
damping term $\Gamma (x)$ is needed. For simplicity we use a
constant dissipation $\Gamma (x) = \gamma$ in our numerical
simulation. Thus the explicit expression for the escape rate for
constant dissipation and arbitrary $g_1 (x)$ and $g_2 (x)$ becomes
\begin{eqnarray}\label{eq49}
k & = & \frac{\omega_0}{2\pi \omega_b} \left [ \frac{\gamma k_BT +
g^2 (x_b)}{\gamma k_BT + g^2 (x_0)} \right ]^{1/2}
\left[\left({\frac{\gamma}{2}}^2 + \omega_b^2
\right)^{1/2} -\frac{\gamma}{2} \right ] \nonumber \\
&& \times \exp \left ( - \frac{\gamma E}{\gamma k_BT + g^2 (x_b)}
\right ),
\end{eqnarray}

\noindent where the quantities $g (x_0)$ and $g (x_b)$ are evaluated
around the potential minima ($x \approx x_0$) and maxima ($x \approx
x_b$), respectively, using the function $g(x)$ defined in
Eq.(\ref{eq26}). In the above equation the quantity
$g^2(x_b)/\gamma$ in the exponential factor together with the
thermal energy $k_BT$ constitutes the effective temperature, a
typical signature of the open system.
\cite{popov,jrc1,jrc2,jrc3,jrc4,jrc5} Another effect of space
dependent diffusion in the dynamics is the presence of reaction
coordinate ($x_0$ and $x_b$) in the steady state rate expression
(\ref{eq49}) which is typically absent in the standard Kramers'
expression \cite{ref1} (see Eq.(\ref{eq52}) below). For
$D_\epsilon=0$ and $D_\pi=0$, i.e., in the absence of external
driving, Eq.(\ref{eq49}) yields Kramers' rate expression for pure
thermal fluctuations valid in the moderate to strong damping limit
\cite{ref1}
\begin{eqnarray}\label{eq52}
k_{Kramers} & = & \frac{\omega_0}{2 \pi \omega_b}
\left[\left({\frac{\gamma}{2}}^2 + \omega_b^2 \right)^{1/2}
-\frac{\gamma}{2} \right ] \nonumber \\
&& \times \exp \left(- \frac{E}{k_B T}\right),
\end{eqnarray}

\noindent which can be also verified easily using the explicit forms
of the parameters $D_0$, $D_b$ and $\Lambda$ in Eq.(\ref{eq48}) in
the limit $\epsilon (t) = \pi (t) = 0$. It is thus clear that in the
absence of the external noise processes, the derived rate expression
(\ref{eq49}) reduces to standard Kramers' escape rate in the
moderate to large damping regime.

\begin{figure}[t]
\includegraphics[width=0.7\columnwidth,angle=-90]{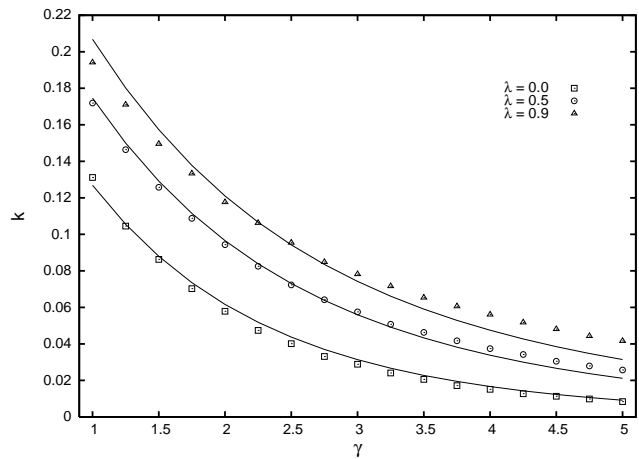}
\caption{Plot of barrier crossing rate $k$ as a function of
dissipation constant $\gamma$ for various values of correlation
parameter $\lambda$ and for $g_1 (x)=x$ and $g_2(x)=1$. The solid
lines are drawn from the theoretical expression, Eq.(\ref{eq49}) and
the symbols are the results of numerical simulation of
Eq.(\ref{eq1}). The values of the parameters used are $k_BT = 0.1$,
$D_\epsilon = D_\pi = 0.1$ and $\lambda$ = 0, 0.5 and 0.9.}
\end{figure}

To study the dynamics, we consider a model cubic potential of the
form $V(x) = b_1x^2 - b_2 x^3$ where, $b_1$ and $b_2$ are the two
constant parameters with $b_1, b_2 > 0$, so that the potential
barrier height becomes $4 b_1^3/27 b_2^2$ and $x_b = 2b_2/3b_1$. In
our numerical simulation we have used $b_1=b_2=1$. While numerically
solving the Langevin equation (\ref{eq1}) for constant dissipation,
$\gamma$, we used a specific combination of the multiplicative
terms, e.g., $g_1 (x) = x$ and $g_2 (x) = 1$. Although both the
external noise processes in our model are multiplicative in nature,
the specific choice of the external noise processes we adopt in the
numerical simulation captures the essential feature of our model
(see the discussion in the following paragraph). We then numerically
solve the Langevin equation (\ref{eq1}) by employing stochastic
simulation algorithm. \cite{jms} The numerical rate has been defined
as the inverse of the mean first passage time
\cite{jrc1,jrc3,jrc4,tgv} and has been calculated by averaging over
10,000 trajectories. To ensure the stability of our simulation, we
have used a small integration time step $\Delta t=0.001$.

In Fig.~1, we show the comparison between the simulation result and
theoretical expression where the escape rate $k$ is plotted as a
function of the dissipation constant, $\gamma$, in the moderate to
large damping regime where our theory is valid, for various values
of the degree of correlation $\lambda$ and observe that for a given
$\gamma$, the escape rate $k$ increases with an increase in
$\lambda$. The result shows a reasonable agreement between the
theory and the simulation. It is easy to check that for a particular
combination of $g_1(x)$ and $g_2(x)$, by increasing the correlation
parameter $\lambda$ one basically increases the value of the
function $g(x)$ (see Eq.(\ref{eq26})) evaluated at $x \approx x_b$.
This increment in $g(x_b)$ increases the effective temperature like
quantity $k_BT+(g^2(x_b)/\gamma)$ for a fixed value of $\gamma$
which in turn pumps more energy into the system which eventually
helps in crossing the barrier and increases the escape rate. Another
way to explain this phenomenon is to look at the diffusion
coefficient, $A_b$ across the barrier. From the expression of $A(x)$
(see Eq.(\ref{eq25})) evaluated at $x \approx x_b$, i.e., $A_b =
\gamma k_BT + g^2 (x_b)$ it is clear that $g (x_b)$ increases with
an increase in $\lambda$ for a particular choice of $g_1(x)$ and
$g_2(x)$ and for a fixed value of $\gamma k_BT$. This increment in
the diffusion across the barrier enhances the reaction rate as
observed. This is the central result of this paper.


\section{Conclusions}

In conclusion, we have extended our recently developed theoretical
approach of studying escape rate from a metastable state under the
influence of external additive cross-correlated noise processes
\cite{jrc5} to investigate the effect of multiplicative noise. In
contrast to our previous approach of using effective correlated
noise constructed from two additive colored noise processes (see
Eq.(1) of Ref.~38), the correlated noise used in the present work
has been constructed from multiplicative white noises. The
multiplicative nature of the correlated noises introduces a space
dependent diffusion in the resultant Fokker-Planck equation
(\ref{eq24}) and modifies the exponential as well as the
pre-exponential factor of the steady state rate expression
(\ref{eq49}). To check the validity of the steady state analytical
rate expression, we have performed numerical simulation of the
starting Langevin equation (\ref{eq1}) with one multiplicative noise
and the other additive one, which shows a satisfactory agreement
between the theory and the numerics. The analytical development as
well as the corresponding numerical simulation lead us to conclude
that the enhancement of rate is possible by increasing the degree of
correlation of the external fluctuations. So far, we have used the
formalism of Markovian stochastic processes in this paper which can
be extended to look into the dynamics within the framework of
non-Markovian formalism. We plan to address this issue in our future
communication.

\begin{acknowledgments}
JRC and SC would like to acknowledge the UGC, India for funding
through the schemes PSW-103/06-07 (ERO) and 32-304/2006 (SR).
SKB acknowledges support from Department of Physics, Virginia Tech.
\end{acknowledgments}


\appendix


\section{Derivation of the Fokker-Planck equation}

In this appendix we give the detailed calculation of constructing
the Fokker-Planck equation (\ref{eq24}) with space dependent
diffusion corresponding to equation (\ref{eq1}) which can be written
as:
\begin{eqnarray}\nonumber
\dot{x} &=& v, \\
\dot{v} &=& -\Gamma(x) v - V^{\prime}(x) + h(x)f(t) \\ \nonumber & +
& g_1(x) \epsilon (t) + g_2(x) \pi(t),
\end{eqnarray}

\noindent We then rewrite the above equation in the following form:
\begin{eqnarray}\nonumber
\dot{u_1}=F_1(u_1,u_2,t;f(t),\epsilon(t),\pi(t)),
\\\label{eq6}
\dot{u_2}=F_2(u_1,u_2,t;f(t),\epsilon(t),\pi(t)),
\end{eqnarray}

\noindent where we use the following abbreviations
\begin{eqnarray}
\label{eq7}
u_1 = x, u_2 & = & v, \\
\label{eq 8} F_1 = v , F_2 & = &- \Gamma(x)v -
V^{\prime}(x)+h(x)f(t)+g_1(x)\epsilon(t)\nonumber \\
&& +g_2(x) \pi(t).
\end{eqnarray}

\noindent The vector $u$ with components $u_1$ and $u_2$ thus
represents a point in a two-dimensional `phase space' and equation
(\ref{eq6}) determines the velocity at each point in the phase
space. The conservation of the phase points now asserts the
following linear equations of motion for density $\rho(u,t)$ in
\textit{phase space}. \cite{ref14}
\begin{equation}\label{eq9}
\frac{\partial}{\partial
t}\rho(u,t)=-\sum_{n=1}^2\frac{\partial}{\partial
u_n}F_n(u,t,f(t),\epsilon(t),\pi(t))\rho(u,t),
\end{equation}

\noindent or more compactly
\begin{equation}\label{eq10}
\frac{\partial \rho}{\partial t}=-\nabla\cdot F \rho.
\end{equation}

\noindent Our next task is to find out a differential equation whose
average solution is given by $\langle \rho \rangle$, \cite{ref14}
where the stochastic averaging has to be performed over both the
internal noise process $f(t)$ and the external noise processes
$\epsilon(t)$ and $\pi(t)$. To this end we note that $\nabla\cdot F$
can be partitioned into two parts: a constant part $\nabla \cdot
F_0$ and a fluctuating part $\nabla \cdot F_1(t)$ containing these
fluctuations. Thus, we write
\begin{eqnarray}\label{eq11}
&& \nabla\cdot F(u,t,f(t),\epsilon(t),\pi(t)) \nonumber \\
&& =  \nabla \cdot F_0(u) \nonumber +\beta \nabla\cdot
F_1(u,t,f(t),\epsilon(t),\pi(t)),
\end{eqnarray}

\noindent where $\beta$ is a parameter (we put it as an external
parameter to keep track of the perturbation equation; we put
$\beta=1$ at the end of the calculation) and also note that $\langle
F_1(t) \rangle =0$. Equation (\ref{eq10}) therefore takes the
following form:
\begin{equation}\label{eq12}
\dot{\rho}(u,t)=(A_0+\beta A_1)\rho(u,t),
\end{equation}

\noindent where $A_0= - \nabla \cdot F_0$ and $A_1=\nabla \cdot
F_1$. The symbol $\nabla$ is used for the operator that
differentiates everything that comes after it with respect to $u$.
Making use of one of the main results for the theory of linear
equation of the form (\ref{eq12}) with multiplicative noise,
\cite{ref14} we derive an average equation for $\rho$ [ $\langle
\rho \rangle = P(u,t)$ the probability density of $u(t)$],
\begin{eqnarray}\label{eq13}
\frac{\partial P}{\partial t} & = & \left\{
A_0+\beta^2\int_0^{\infty}d\tau \langle A_1(t)\exp(\tau
A_0)A_1(t-\tau) \rangle \right. \nonumber \\
&& \left. \times \exp{(-\tau A_0)} \right \}P.
\end{eqnarray}

\noindent Equation (\ref{eq13}) is exact when the correlation times
of fluctuations tend to zero. Using the expressions for $A_0$ and
$A_1$ we obtain
\begin{eqnarray}\label{eq14}
\frac{\partial P}{\partial t} & = & \left\{ -\nabla\cdot
F_0+\beta^2\int_0^{\infty}d\tau \left \langle \nabla\cdot
F_1(t) \right. \right. \nonumber \\
&& \left. \left. \times \exp(-\tau\nabla\cdot F_0)
\nabla\cdot F_1(t-\tau) \right
\rangle \exp(\tau \nabla\cdot F_0) \right\}P . \nonumber \\
\end{eqnarray}

\noindent The operator $\exp(\tau \nabla \cdot F_0)$  in the above
equation provides the solution to the equation
\begin{equation}\label{eq15}
\frac{\partial y(u,t)}{\partial t}=-\nabla \cdot F_0 y(u,t),
\end{equation}

\noindent ($y$ signifies the unperturbed part of $\rho$), which can
be found explicitly in terms of characteristic curves. The equation
\begin{equation}\label{eq16}
\dot{u}=F_0(u),
\end{equation}

\noindent for fixed $t$ determines a mapping from $u(\tau=0)$ to
$u(\tau)$ \textit{i.e.}, $u\rightarrow u^{\tau}$ with the inverse
$(u^\tau)^{-\tau}=u$. The solution of equation (\ref{eq15}) is
\begin{equation}\label{eq17}
y(u,t)=y(u^{-t},0) \left |\frac{d(u^{-t})}{d(u)} \right
|=\exp(-t\nabla\cdot F_0)y(u,0) ,
\end{equation}

\noindent $|d(u^{-t})/d(u)|$ being a Jacobian determinant. The
effect of $\exp(-t\nabla\cdot F_0)$ on $y(u)$ is as follows:
\begin{equation}\label{eq18}
\exp(-t\nabla\cdot F_0) y(u,0)=y(u^{-t},0) \left
|\frac{d(u^{-t})}{d(u)} \right |.
\end{equation}

\noindent The above simplification when put in equation (\ref{eq14})
yields
\begin{eqnarray}\label{eq19}
\frac{\partial P(u,t)}{\partial t} & = & \nabla \cdot \left\{
-F_0+\beta^2 \int_0^\infty d\tau \left | \frac{d(u^{-\tau})}{d(u)}
\right | \right. \nonumber \\
&& \times \left \langle F_1(u,t) \nabla_{-\tau}\cdot
F_1(u^{-\tau},t-\tau) \right \rangle \nonumber \\
&& \left. \times \left |
\frac{d(u)}{d(u^{-\tau})} \right | \right\} P(u,t).
\end{eqnarray}

\noindent where $\nabla_{-\tau}$ denotes differentiation with
respect to $u^{-\tau}$. We put $\beta=1$ for the rest of the
treatment. Identifying $u_1=x$ and $u_2=v$ we now write
\begin{eqnarray}
F_{01} & = & v, F_{11}=0, \nonumber \\
F_{02} & = & -\Gamma(x)v- V^{\prime}(x), \nonumber \\
F_{12} & = & h(x)f(t) + g_1(x) \epsilon(t) + g_2(x) \pi(t).
\end{eqnarray}

\noindent In this situation, equation (\ref{eq19}) now reduces to
\begin{eqnarray}\label{eq21}
\frac{\partial P}{\partial t}& = &-\frac{\partial}{\partial x}(v
P)+\frac{\partial}{\partial v} \left\{\Gamma(x)v+V'(x) \right\}P\nonumber\\
& & + \frac{\partial}{\partial v}\int_0^\infty d\tau \left
\langle[h(x) f(t)+g_1(x)\epsilon(t)+g_2(x)\pi(t) \right.
\nonumber \\
&& + \frac{\partial}{\partial v^{-\tau}}
\{h(x^{-\tau})f(t-\tau)+g_1(x^{-\tau})\epsilon(t-\tau)
\nonumber \\
&& \left. + g_2(x^{-\tau})\pi(t-\tau))\}] \right \rangle P,
\end{eqnarray}

\noindent where we have used the fact that the Jacobian obeys the
equation
\begin{equation}\nonumber
\frac{d}{dt}\log
\left|\frac{d(x^t,v^t)}{d(x,v)}\right|=\frac{\partial v}{\partial
x}+\frac{\partial}{\partial v}\left\{-\Gamma v+V'(x)\right\} =
-\Gamma,
\end{equation}

\noindent so that the Jacobian becomes $\exp(-\Gamma(x)t)$. Now
neglecting the terms $O(\tau^2)$ we may have $x^{-\tau} = x - \tau
v$ and $v^{-\tau} = v + \Gamma \tau v + \tau V(x)$. The above two
equations yield
\begin{equation}\label{eq22}
\frac{\partial}{\partial v^{-\tau}}=(1-\Gamma
\tau)\frac{\partial}{\partial v}+\tau \frac{\partial}{\partial x}.
\end{equation}

\noindent Taking this in to consideration of equation (\ref{eq22}),
equation (\ref{eq21}) can be simplified in the following form:
\begin{eqnarray}\label{eq23}
\frac{\partial P}{\partial t} &=&  -\frac{\partial (v P)}{\partial x
}
+\frac{\partial}{\partial v} [\Gamma(x)v+V'(x)]P \nonumber \\
&& + \frac{\partial^2}{\partial v^2} [k_BT \Gamma(x) + g^2(x)]P,
\end{eqnarray}

\noindent where
\begin{eqnarray}
g(x) & = & \left \{ D_\epsilon g_1^2(x) + 2 \lambda \sqrt{D_\epsilon
D_\pi} g_1(x) g_2(x) \right. \nonumber \\
&& \left. + D_\pi g_2^2(x) \right \}^{1/2}.
\end{eqnarray}


\noindent Defining
\begin{equation}
A(x) = k_B T \Gamma(x) + g^2(x),
\end{equation}

\noindent the above equation (\ref{eq23}) can be written as
\begin{eqnarray}
\frac{\partial P}{\partial t}& = & -v\frac{\partial P}{\partial
x}+[\Gamma(x)v+V'(x)]\frac{\partial P}{\partial v} +
A (x) \frac{\partial^2 P}{\partial v^2} \nonumber \\
&& + \Gamma(x)P,
\end{eqnarray}

\noindent
which is our required Fokker-Planck equation.


\section{Calculation of the escape rate}

In this appendix we show the detailed calculation of the escape rate
for nonequilibrium open system. The technique we adopt here
resembles our previous approaches \cite{jrc1,jrc2,jrc5} but makes
the current paper self contained.

Following Kramers, \cite{ref1} we make ansatz that the
non-equilibrium steady state probability $P_b^{st} (x,v)$ generating
a non-vanishing diffusion current across the barrier is given by
\begin{equation}
P_b^{st} (x,v)= \exp \left ( - \frac{v^2}{2 D_b} - \frac{V(x)}{D_b}
\right ) G(x,v), \label{eq32}
\end{equation}

\noindent where $D_b=A_b / \Gamma(x_b)$. Inserting Eq.(\ref{eq32})
in (\ref{eq31}), we obtain the equation for $G(x,v)$ using the
steady state in the neighborhood of $x_b$
\begin{equation}
-v\frac{\partial G}{\partial x} -[\omega_b^2 (x - x_b) + \Gamma (x)
v] \frac{\partial G}{\partial v} + A_b \frac{\partial^2 G}{\partial
v^2} = 0.   \label{eq33}
\end{equation}

\noindent At this point we set
\begin{equation}\label{eq34}
y=v+a(x-x_b),
\end{equation}

\noindent where $a$ is a constant to be determined. With the help of
the transformation (\ref{eq34}), Eq.(\ref{eq33}) reduces to
\begin{equation}\label{eq35}
A_b \frac{d^2 G}{dy^2} - [\omega_b^2 (x - x_b) + \{ \Gamma
(x_b) +a \} v] \frac{dG}{dy} = 0.
\end{equation}

\noindent Now let
\begin{equation}\label{eq36}
[\omega_b^2 (x - x_b) + \{ \Gamma (x_b) +a \} v] = - \mu y,
\end{equation}

\noindent where $\mu$ is another constant. By virtue of the relation
(\ref{eq36}), Eq.(\ref{eq35}) becomes
\begin{equation}\label{eq37}
\frac{d^2 G}{dy^2} + \Lambda y \frac{dG}{dy} = 0,
\end{equation}

\noindent where $\Lambda = \mu /A_b$ with $A_b= k_B T \Gamma(x_b) +
g^2(x_b)$. The constant $\mu$ and $a$ must satisfy the following
equations simultaneously:
\begin{equation}
-\mu a = \omega_b^2 \; {\rm and} \;  -\mu = \Gamma(x_b)+ a.
\label{eq37a}
\end{equation}

\noindent This implies that the constant $a$ must satisfy the
following quadratic equation
\begin{equation}\nonumber
a^2 + \Gamma(x_b) a - \omega_b^2 = 0,
\end{equation}

\noindent which allows the solution for $a$ as
\begin{equation}\label{eq39}
a_{\pm} = \frac {1}{2} \left \{ - \Gamma (x_b) \pm \sqrt{\Gamma^2
(x_b) + 4 \omega_b^2 } \right \}.
\end{equation}

\noindent Thus, the general solution of (\ref{eq37}) is
\begin{equation}\label{eq40}
G(y) = G_2 \int_0^y \exp \left(-\frac{\Lambda z^2}{2} \right)dz +
G_1,
\end{equation}

\noindent where $G_1$ and $G_2$ are two constants of integration. We
look for a solution which vanishes for large $x$. This condition is
satisfied if the integration in (\ref{eq40}) remains finite for $|y|
\rightarrow + \infty $. This implies that $\Lambda
> 0$ so that only $a_{-}$ becomes relevant. Then the requirement
$P_{b}(x,v)\rightarrow 0$ for $x\rightarrow +\infty$ yields
\begin{equation}\label{eq41}
G_1 = G_2 \sqrt{\frac{\pi}{2\Lambda}}.
\end{equation}

\noindent Thus we have
\begin{equation}\label{eq42}
G(y) = G_2 \left [ \sqrt{\frac{\pi}{2\Lambda}} + \int_0^y \exp \left
( -\frac{\Lambda z^2}{2} \right ) dz \right ],
\end{equation}

\noindent and correspondingly,
\begin{eqnarray}\label{eq43}
P_b^{st} (x,v) &=& G_2 \left [ \sqrt{\frac{\pi}{2\Lambda}} +
\int_0^y \exp
\left (- \frac{\Lambda z^2}{2} \right )dz \right ] \nonumber \\
&& \times \exp \left (-\frac{v^2}{2D_b} - \frac{V(x)}{D_b} \right ).
\end{eqnarray}

\noindent The current across the barrier associated with this steady
state distribution is given by
\begin{equation}\nonumber
j = \int_{-\infty}^{+\infty} v P_b^{st} (x \approx x_b, v) dv,
\end{equation}

\noindent which may be evaluated using (\ref{eq43}) and the
linearized version of $V(x)$, namely $V(x) \approx E_b - \omega_b^2
(x-x_b)^2/2$ as
\begin{equation}\label{eq44}
j = G_2 \sqrt {\frac{2 \pi}{\Lambda + D_b^{-1}}} D_b \exp
\left(-\frac{E_b}{D_b} \right).
\end{equation}

\noindent To determine the remaining constant $G_2$ we note that as
$x\rightarrow -\infty$, the pre-exponential factor in (\ref{eq43})
reduces to $ G_2 \sqrt{2 \pi /\Lambda}$. We then obtain the reduced
distribution function in $x$ as
\begin{equation}\label{eq45}
{\tilde P}_b^{st}(x\rightarrow -\infty) = 2 \pi G_2 {\sqrt
\frac{D_b}{\Lambda}} \exp \left(-\frac{V(x)}{D_b} \right),
\end{equation}

\noindent where we have used the definition for the reduced
distribution \cite{jrc1,jrc2,jrc5} as
\begin{equation}\nonumber
{\tilde P}(x) = \int^{-\infty}_{+\infty} P(x,v) dv.
\end{equation}

\noindent Similarly, we derive the reduced distribution in the left
well around $x \approx x_0$ using Eq.(\ref{eq28}) where the
linearized potential is $V(x) \approx E_0 + \omega_0^2 (x-x_0)^2/2$,
\begin{equation}\label{eq46}
{\tilde P}_0^{st}(x) = \frac{1}{Z} \sqrt {2\pi D_0} \exp
\left(-\frac {\omega_0^2 (x-x_0)^2}{2D_0}\right),
\end{equation}

\noindent with the normalization constant given by $1/Z =
\omega_0/(2 \pi D_0)$ The comparison of the distribution
(\ref{eq45}) and (\ref{eq46}) near $x\approx x_0$, gives,
\begin{equation}\label{eq47}
G_2 = \sqrt {\frac {\Lambda}{D_b}} \frac {\omega_0}{2\pi \sqrt {2
\pi D_0}}.
\end{equation}

\noindent Hence, from (\ref{eq44}), the normalized current or the
barrier crossing rate $k$, for moderate to large friction regime is
given by
\begin{equation}
k = \frac {\omega_0}{2\pi}
\frac{D_b}{\sqrt{D_0}}\sqrt{\frac{\Lambda}{1+ \Lambda D_b}} \exp
\left(-\frac{E}{D_b} \right),
\end{equation}
\noindent where $E=E_b-E_0$ is the potential barrier height.


\end{document}